\documentstyle[11pt]{article}

\newcommand{\nom}[2]{\left[ \begin{array}{c} #1 \\ #2 \end{array} \right]}

\begin{document}

\title{Multinomials and
Polynomial Bosonic Forms for the Branching Functions of the 
$\widehat{su}(2)_{M}\times \widehat{su}(2)_{N}/\widehat{su}(2)_{M+N}$
Conformal Coset Models}
\author{Anne Schilling \thanks{e-mail address: anne@insti.physics.sunysb.edu}\\
\footnotesize{{\em Institute for Theoretical Physics,}}\\ 
\footnotesize{{\em State University of New York at Stony Brook, Stony Brook, 
NY 11794-3800}}}

\date{}

\maketitle 

\begin{abstract}
We give explicit expressions for the q-multinomial generalizations of the
q-binomials and Andrews' and Baxter's q-trinomials. We show that the 
configuration sums for the generalized RSOS models in regime III studied by 
Date et al. can be expressed in terms of these multinomials. This generalizes 
the work of ABF and AB where configuration sums of statistical mechanical 
models have been expressed in terms of binomial and trinomial coefficients.
These RSOS configuration sums yield the branching functions
for the $\widehat{su}(2)_{M}\times \widehat{su}(2)_{N}/\widehat{su}(2)_{M+N}$
coset models. The representation in terms of multinomials gives Rocha-Caridi 
like formulas whereas the representation of Date et al. gives a double sum 
representation for the branching functions.
\end{abstract}

\section{Introduction}
\label{sec-intro}
\setcounter{equation}{0}

There are many polynomial generalizations of conformal field theory 
characters/branching functions \cite{ABF}-\cite{BM1}.
Often, these polynomials come from finite dimensional statistical mechanical
configuration sums and yield the characters of conformal field theories
as the order of the polynomial tends to infinity. Polynomial expressions
are useful since they often permit derivations of recursion relations in the
order of the polynomial. These recursion relations allow the proof of
nontrivial identities between polynomial expressions and hence also for the
corresponding infinite series obtained by taking the order of the polynomial
to infinity.

In their bosonic form most of these polynomial representations use 
$q$-binomials  or Andrews' and Baxter's \cite{AB} $q$-trinomials. In 
particular, Andrews, Baxter and Forrester \cite{ABF} express
the configuration sums of their ABF model in terms of binomials which are
polynomial forms of the branching functions of the $M(p,p+1)$, $M(2,2p+1)$,
$M(2p-1,2p+1)$ minimal models and the $Z_p$ parafermion model in the different
regimes. These bosonic polynomial forms have been generalized to the arbitrary
minimal models $M(p,p')$ by Forrester and Baxter \cite{FB}. Further, Andrews 
and Baxter \cite{AB} give bosonic polynomial versions of the branching 
functions of the superconformal unitary models $SM(5,7)$ and the minimal 
models $M(2,7)$ in terms of trinomials. Warnaar and Pearce \cite{WP}, 
\cite{WP1} use trinomials to express $E_{6,7,8}$ type Rogers-Ramanujan 
identities. Berkovich, McCoy and Orrick \cite{BMO} give polynomial forms for 
the $SM(2,4\nu)$ and $M(2\nu-1,4\nu)$ characters in terms of trinomials. These
characters have been further investigated by Berkovich and McCoy \cite{BM1}.

In this paper we generalize the $q$-binomial and Andrews' and Baxter's 
$q$-trinomial coefficients \cite{AB} to $q$-multinomial coefficients for
which we give explicit formulas in equation~(\ref{multi}). We further write 
the configuration sums for the generalized RSOS models in regime III studied 
by Date et al. \cite{Kyoto}-\cite{Date3} in terms of these multinomials (see 
equation~(\ref{X_multi}) and (\ref{X_multi1})). The configuration sums of the
generalized RSOS models in regime III which are polynomials coincide
with the branching functions of the
$\widehat{su}(2)_{M}\times \widehat{su}(2)_{N}/\widehat{su}(2)_{M+N}$ coset
models as the order of the polynomial limits to infinity.

Date et al. \cite{Date1}, \cite{Date2} give several representations for these 
RSOS configuration sums: 1) double sum expressions, 2) single sums in terms of
binomials and 3) for a special case an expression in terms of multinomials 
previously defined by Andrews \cite{Andrews2}.
These multinomials, however, differ from the ones used in this paper and
are not appropriate for taking the limit that gives the branching functions.

The RSOS models studied by Date et al. generalize the ABF \cite{ABF} and AB 
\cite{AB} statistical models which correspond to the
cases $N=1$ and $N=2$, respectively. ABF and AB find expressions 
for the configuration sums in terms of bi- and trinomials. These configuration
sums yield Rocha-Caridi type \cite{RC} expressions for the branching functions
whereas Date et al. \cite{Date1}, \cite{Date2} obtain double sum expressions 
which, in turn, can be expressed in terms of elliptic 
theta functions. In this paper we generalize the expressions given by ABF and 
AB and write the configuration sums in terms of multinomials which generalize
AB's trinomials \cite{AB}. They yield
branching functions of the type given in \cite{KB}.

For the proof of the configuration sums in terms of multinomials one can use 
the fact that the configuration sums of the RSOS models obey certain recursion
relations. We will show that this induces certain depth one recursion relations
for the multinomials (see equation (\ref{multi_rec})) which are proven in
section \ref{sec_proof}.

Usually one finds two very distinct solutions to these kinds of 
recursion relations, a fermionic solution and a bosonic solution. By computing
both one gets Rogers-Ramanujan type identities. Schur \cite{Schur} was 
the first to prove such Bose-Fermi identities by proving the 
Rogers-Ramanujan identities via recursion relations.
Recently fermionic type solutions of the RSOS models and hence the
$\widehat{su}(2)_{M}\times \widehat{su}(2)_{N}/\widehat{su}(2)_{M+N}$ coset
models have been found and proven. The characters for the
case $N=1$ and also the identity character for arbitrary $N$ have been 
conjectured by Kedem et al. \cite{Kedem}. A polynomial form of these
characters has been conjectured by Melzer \cite{Melzer} and proven
by Berkovich \cite{Berkovich} and Warnaar \cite{Warnaar1}, \cite{Warnaar2}.
The fermionic branching functions for $N=2$ have been conjectured by
Baver and Gepner \cite{BG}. Polynomial fermionic forms for general $N$
have recently been found and proven by the author \cite{me}.

As the order of the polynomial configuration sum in terms of multinomials
tends to infinity it factorizes into a Rocha-Caridi type sum and a 
parafermionic configuration sum. Therefore for the structure of the
branching function of the
$\widehat{su}(2)_{p-N-2}\times \widehat{su}(2)_{N}/\widehat{su}(2)_{p-2}$
coset models we obtain \newpage
\begin{eqnarray}
\label{form}
c_{r,s}^{(l)}&\sim &\frac{1}{(q)_{\infty}}
 \sum_{m=0}^{N/2} {\rm (parafermion\; piece)}(m,l)\nonumber\\
&&\times
 \left\{\sum_{j\in {\bf Z},m_{r,s}(j)\equiv m} q^{\frac{j}{N}(jpp'+pr-p's)}
 -\sum_{j\in {\bf Z},m_{r,-s}(j)\equiv m} q^{\frac{1}{N}(pj+s)(p'j+r)}\right\}
\end{eqnarray}
where the first sum runs over $m$ integer for $l$ odd and $m$ half-integer 
for $l$ even, $m_{r,s}(j)=|\frac{r-s}{2}+pj \bmod N|$ (where the convention
$N/2<(x \bmod N)\leq N/2$ is used) and $p'=p-N$. The term in 
brackets is of the Rocha-Caridi form. Details are given in 
section~\ref{sec_partition}.

The bosonic polynomial forms of the 
$\widehat{su}(2)_1\times \widehat{su}(2)_{p-3}/\widehat{su}(2)_{p-2}$
coset models as given by ABF \cite{ABF} have been generalized by Forrester 
and Baxter \cite{FB} to expressions for coset models 
$\widehat{su}(2)_1\times\widehat{su}(2)_{m}/\widehat{su}(2)_{m+1}$ with
fractional levels $m=\frac{p'}{p-p'}-2$ or $m=-\frac{p}{p-p'}-2$. 
Fermionic polynomial forms of these models have been given by Berkovich and
McCoy \cite{BM}. Hence one may speculate that multinomials will also turn out 
to be useful for coset models
$\widehat{su}(2)_N\times\widehat{su}(2)_M/\widehat{su}(2)_{N+M}$ where 
$M$ is fractional. An example of this has already been seen in \cite{Ole}.

The characters/branching functions of conformal models can also be obtained
from representations of the Virasoro algebra directly by the Feigin and 
Fuchs \cite{FF} construction which uses cohomological methods to mod out
singular vectors. The Rocha-Caridi character formula has been derived this 
way \cite{RC} and gives the characters for the coset models
$\widehat{su}(2)_{M}\times \widehat{su}(2)_{N}/\widehat{su}(2)_{M+N}$
for $N=1$. Similarly, the branching functions and characters for $N=2$ have 
been derived in \cite{GKO} and for arbitrary $N$ in \cite{KB} which 
lead to products of Rocha-Caridi type expressions times string functions. 
The parafermionic piece in (\ref{form}) can be expressed in terms of the 
$\widehat{su}(2)_N$ string functions which allows us to make a connection 
between our results and the work in \cite{KB}.

This paper is structured as follows. In section \ref{sec_multi}
we review the binomial and trinomial coefficients, give multinomial
generalizations of these and state some properties of the multinomials. In
section \ref{sec_partition} the representation of the RSOS configuration sums
in terms of multinomials is derived by using the path space interpretation
of the RSOS models which yield recursion relations for the configuration sums.
The results are compared with the representations previously obtained by Date
et al.. The branching functions of the
$\widehat{su}(2)_{M}\times \widehat{su}(2)_{N}/\widehat{su}(2)_{M+N}$ coset
models are obtained by taking the order of the configuration sum polynomials
to infinity and are compared with the results in \cite{KB}.
In section \ref{sec_proof} we prove the recursion relations of the 
multinomials which are essential for the proof of the configuration sum
formulas in section \ref{sec_partition}. We conclude in section \ref{summary}
with a summary and discussion of open questions.

\section{Multinomial coefficients}
\label{sec_multi}
\setcounter{equation}{0}

We start our discussion of the multinomials by briefly reviewing the 
$q$-binomial and $q$-trinomial coefficients. To this end let us define
\begin{equation}
(q)_n=\prod_{i=1}^{n}(1-q^i)
\end{equation}
The $q$-binomials -also called Gau{\ss}ian polynomials- are then defined as
\begin{equation}
\label{Gauss}
\nom{L}{r}_1=\left\{ \begin{array}{ll} \frac{(q)_L}{(q)_r(q)_{L-r}} &
 {\rm if} \, 0\leq r\leq L\\ 0 & {\rm otherwise} \end{array}\right. .
\end{equation}
Andrews and Baxter \cite{AB} define $q$-trinomials as
\begin{equation}
\label{tri}
\nom{L}{r}_2^{(n)}=\sum_{m\geq 0}\frac{q^{m(m+r-n)}(q)_L}{(q)_m(q)_{m+r}
 (q)_{L-2m-r}}
\end{equation}
where $n=0,1$ and $r=-L,-L+1,\ldots,L$. They further define
\begin{equation}
\label{T2}
T_n(L,r)=q^{\frac{L(L-n)-r(r-n)}{2}}\nom{L}{r}_{q^{-1},2}.
\end{equation}
(Notice that the $T_n$ defined in (\ref{T2}) is actually 
$T_n(L,r)=T_n(L,r,q^{\frac{1}{2}})$ in Andrews' and Baxter's notation). 
In \cite{BMO} Berkovich, McCoy and Orrick generalize the trinomials
$\nom{L}{r}_2^{(n)}$ and also $T_n(L,r)$ to arbitrary $n\in {\bf Z}$ to write
the bosonic characters for the $SM(2,4\nu)$ models. We will however only
generalize Andrews' and Baxter's trinomial coefficient.

Using
\begin{equation}
\label{q_inverse}
(q^{-1})_n=(-)^nq^{-\frac{n(n+1)}{2}}(q)_n
\end{equation}
and making the variable change $m\rightarrow\frac{L}{2}-\frac{r}{2}
-\frac{m}{2}$ one derives from (\ref{T2})
\begin{equation}
\label{trinomial}
T_n(L,r)=\sum_{m, \frac{L}{2}-\frac{r}{2}-\frac{m}{2}\in {\bf Z}_{\geq 0}}
 q^{\frac{m^2}{2}-\frac{nm}{2}}\frac{(q)_L}{(q)_{\frac{L}{2}-\frac{r}{2}
 -\frac{m}{2}}(q)_{\frac{L}{2}+\frac{r}{2}-\frac{m}{2}}(q)_m}.
\end{equation}
The sum runs over all $m\in{\bf Z}_{\geq 0}$ such that
$\frac{L}{2}-\frac{r}{2}-\frac{m}{2}\in {\bf Z}_{\geq 0}$ and
$\frac{L}{2}+\frac{r}{2}-\frac{m}{2}\in {\bf Z}_{\geq 0}$.
Notice that Andrews and Baxter write $T_n$ ($n=0,1$) in a different form
\begin{equation}
T_n(L,r,q)=\sum_{j=0}^{L}(-q^n)^j\nom{L}{j}_{1,q^2}\nom{2L-2j}{L-r-j}_{1,q}.
\end{equation}
This form of the trinomials is however harder to generalize.
In \cite{Andrews} Andrews gives a recursive definition for the multinomial
coefficients $\nom{L}{r}_N^{(0)}$, but no explicit formulas and properties
(e.g. recursion relations) are given. The multinomials used by Date et al.
(cf. equation (3.29) in \cite{Date1}) are related to $\nom{L}{r}_N^{(0)}$.

Let us define the $(N+1)$-nomial coefficients which generalize the trinomial
coefficients in (\ref{trinomial}) as
\begin{eqnarray}
\label{multi}
T_n^{(N)}(L,r)\equiv \sum_{m\in {\bf Z}_{\geq 0}^{N-1},
 \frac{L}{2}-\frac{r}{N}-e_1C^{-1}m\in {\bf Z}_{\geq 0}} 
 q^{mC^{-1}m-e_nC^{-1}m}
 \nonumber\\
 \times \frac{(q)_L}{(q)_{\frac{L}{2}-\frac{r}{N}-e_1C^{-1}m}
 (q)_{\frac{L}{2}+\frac{r}{N}-e_{N-1}C^{-1}m}}
 \prod_{i=1}^{N-1}\frac{1}{(q)_{m_i}}
\label{def_multi}
\end{eqnarray}
where the subscript $(N)$ denotes the $(N+1)$-nomial, $n=0,1,2,\ldots,N-1$ 
gives the type of the multinomial (similar to $T_0$ and $T_1$ for the 
trinomials) and $r=-\frac{NL}{2},-\frac{NL}{2}+1,\ldots,\frac{NL}{2}$. 
For other values of $r$ the multinomials are defined to be zero. The sum runs 
over all $m=(m_1,m_2,\ldots,m_{N-1})\in {\bf Z}_{\geq 0}^{N-1}$ such that
$\frac{L}{2}-\frac{r}{N}-e_1C^{-1}m\in {\bf Z}_{\geq 0}$ and 
$\frac{L}{2}+\frac{r}{N}-e_{N-1}C^{-1}m\in {\bf Z}_{\geq 0}$.
The vectors $e_i$ are defined as $(e_i)_j=\delta_{i,j}$ for 
$i\in\{1,2,\ldots,N-1\}$ and $e_i=0$ otherwise. $C^{-1}$ is the inverse 
of the $N-1$ dimensional Cartan matrix $C=2-I$ where the incidence matrix 
$I_{ab}=\delta_{a,b+1}+\delta_{a,b-1}$. An explicit formula for $C^{-1}$ is
given by
\begin{equation}
C^{-1}_{i,j}=
\left\{ \begin{array}{ll}
\frac{1}{N}(N-i)j & {\rm for}\;j\leq i\\
C^{-1}_{j,i} & {\rm for}\; j>i
\end{array} \right.
\end{equation}

We can write the expression for $T^{(N)}_n$ in (\ref{multi}) in a slightly 
more elegant way by introducing
\begin{eqnarray}
\label{new_m0}
m_0=\frac{L}{2}-\frac{r}{N}-e_1C^{-1}m\\
\label{new_mN}
m_N=\frac{L}{2}+\frac{r}{N}-e_{N-1}C^{-1}m.
\end{eqnarray}
Then we may write
\begin{equation}
\label{multi_short}
T_n^{(N)}(L,r)=\widetilde{\sum_{\tilde{m}\in {\bf Z_{\geq 0}^{N+1}}}}
 q^{mC^{-1}m-e_nC^{-1}m}(q)_L\prod_{i=0}^{N}\frac{1}{(q)_{m_i}}
\end{equation}
where $\tilde{m}=(m_0,m_1,\ldots,m_N)$, $m=(m_1,m_2,\ldots,m_{N-1})$ and 
$\widetilde{\sum}$ stands for the sum over all $m_i\in {\bf Z}_{\geq 0}$ 
$(i=0,1,2,\ldots,N)$ such that equations (\ref{new_m0}) and (\ref{new_mN}) 
are satisfied.

Notice that the multinomials have the symmetry
\begin{equation}
\label{symmetry}
T_n^{(N)}(L,-r)=T_{N-n}^{(N)}(L,r)
\end{equation} 
where we define $T_N\equiv T_0$. To see this just relabel the summation 
variable $m_i$ in $T_n^{(N)}(L,-r)$ by $m_{N-i}$. To reobtain the previous 
form for $T^{(N)}_n$ $e_n$ gets replaced by $e_{N-n}$ and (\ref{symmetry}) 
follows. From the definition of the multinomials one can immediately read off
the initial condition
\begin{equation}
\label{multi_initial}
T_n^{(N)}(0,r)=\delta_{r,0}.
\end{equation}
Further -but less trivial to see- the multinomials satisfy the
following recursion relations
\begin{eqnarray}
\label{multi_rec}
T_n^{(N)}(L,r)&=&\sum_{k=0}^{n-1}q^{(n-k)(\frac{L}{2}+\frac{r}{N})}
 T_{N-k}^{(N)}(L-1,\frac{N}{2}+r-k)\nonumber\\
 &+&\sum_{k=n}^{N}q^{(k-n)(\frac{L}{2}-\frac{r}{N})}
 T_{N-k}^{(N)}(L-1,\frac{N}{2}+r-k)
\end{eqnarray}
These recursion relations the proof of which will be given in section
\ref{sec_proof} are essential for the derivation of the RSOS configuration
sums. The recursion relations (\ref{multi_rec}) generalize
the recursion relations for the trinomials (cf. (2.16) and (2.19) 
in \cite{AB}).

Notice that in the limit $L\rightarrow \infty$ the multinomials become
the product of $\frac{1}{(q)_{\infty}}$ times the partition functions for the 
$Z_N$-parafermion model, namely
\begin{eqnarray}
\label{limit}
T^{(N)}_n(r)&\equiv& \lim_{L\rightarrow \infty, L\;{\rm even}}T_n^{(N)}(L,r)
\nonumber\\
&=&\frac{1}{(q)_{\infty}}
 \sum_{m\in {\bf Z}^{N-1}_{\geq 0}, \frac{r}{N}+e_1C^{-1}m\in {\bf Z}}
 q^{mC^{-1}m-e_nC^{-1}m}\prod_{i=1}^{N-1}\frac{1}{(q)_{m_i}}.
\end{eqnarray}

The analogue of (\ref{Gauss}) and (\ref{tri}) is given by
\begin{equation}
\label{def_nom}
\nom{L}{r}_N^{(n)}=q^{\frac{1}{N}(\frac{NL}{2}(\frac{NL}{2}-n)-r(r-n))}
 T_n^{(N)}(L,r)_{q^{-1}}.
\end{equation}
An explicit formula for $\nom{L}{\frac{NL}{2}-a}^{(n)}_N$ where now 
$a=0,1,\ldots,NL$ is given by
\begin{eqnarray}
\label{nom}
\lefteqn{\nom{L}{\frac{NL}{2}-a}_N^{(n)}=}\nonumber\\
&&\!\!\!\!\!\!\!\!\!\sum_{j_1+j_2+\ldots+j_N=a}
 q^{\sum_{k=1}^{N-1}(L-j_k)j_{k+1}-\sum_{k=N-n}^{N-1}j_{k+1}}
 \nom{L}{j_1}\nom{j_1}{j_2}\nom{j_2}{j_3}\ldots\nom{j_{N-1}}{j_N}
\end{eqnarray}
where the sum runs over all $j_1,j_2,\ldots,j_N\in {\bf Z}_{\geq 0}$
such that $j_1+j_2+\ldots+j_N=a$.
One can obtain this explicit formula by first replacing 
$m_i\rightarrow m_{N-i}$ in (\ref{multi_short}), using (\ref{q_inverse}) and
identifying
\begin{eqnarray}
&&j_1=L-m_0=\frac{a}{N}+e_1C^{-1}m\nonumber\\
&&j_k=j_{k-1}-m_{k-1}=\frac{a}{N}+(-e_{k-1}+e_k)C^{-1}m\;\;{\rm for}\;
 1<k<N\nonumber\\
&&j_N=j_{N-1}-m_{N-1}=m_N=\frac{a}{N}-e_{N-1}C^{-1}m.
\end{eqnarray}
Equation (\ref{nom}) has been found independently by Warnaar \cite{Ole}.
In Warnaar's notation $\nom{L}{a}_{N,{\rm Warnaar}}^{(n)}=
\nom{L}{\frac{NL}{2}-a}_N^{(n)}$ of (\ref{nom}).

The list of properties of the multinomials as given above is certainly not
exhaustive. Some further properties such as identities which reduce to 
tautologies as $q\rightarrow 1$ and a partition theoretical interpretation
of the multinomials can be found in \cite{Ole}.

\section{Bosonic RSOS configuration sums}
\label{sec_partition}
\setcounter{equation}{0}

Let us first review the path space for the general RSOS models.
The RSOS model that we consider here is defined on a square lattice
${\cal L}$. To each site $i$ on the lattice one associates a state variable
$l_i$ which can take the values $l_i=1,2,\ldots,p-1$. The local height 
probability, i.e. the probability that a certain state variable takes some
particular value, is given as a two dimensional configuration sum and
can be reduced via the corner transfer matrix method \cite{Baxter} to a 
one dimensional configuration sum over a certain path space.

A path in this path space is a string of state variables $\{l_i|1\leq 
i\leq L+2\}$ such that two adjacent state variables $l_i$ and $l_{i+1}$
are {\em admissible}.
Two adjacent state variables $l_i$ and $l_{i+1}$ are called {\em admissible} 
($l_i \sim l_{i+1}$) if they fulfill the following conditions:
\begin{eqnarray}
\label{admissable}
\label{weak}
l_i-l_{i+1}=-N,-N+2,\ldots,N\\
\label{bound}
l_i+l_{i+1}=N+2,N+4,\ldots,2p-N-2
\end{eqnarray}
where $p$ and $N$ are arbitrary positive integers and $i$ runs 
from $1,2,\ldots,L+2$. The model is frozen however unless $p\geq N+3$. 
In terms of the path space $N$ defines the value for the
highest possible step in the vertical direction, $p$ and $L$ 
define the boundaries of the lattice in the vertical and horizontal
direction, respectively. We call a pair $(l_i,l_{i+1})$ {\em weakly admissible}
if (\ref{weak}) is satisfied, but not necessarily (\ref{bound}).

The one dimensional configuration sum for the RSOS models is given 
by \cite{Date1}, \cite{Date2}
\begin{equation}
X_{L}(a|b,c)=\sum_{l_i}q^{\sum_{j=1}^{L} \frac{j}{4} |l_{j+2}-l_j|}
\end{equation}
where $l_1=a, l_{L+1}=b, l_{L+2}=c$ and $l_1\sim l_2\sim \cdots \sim l_{L+2}$.
$X_L$ is uniquely determined by 1)~the initial condition
\begin{equation}
\label{X_initial}
X_0(a|b,c)=\delta_{a,b}
\end{equation}
and 2) the recursion relation
\begin{equation}
\label{X_rec}
X_L(a|b,c)=\sum_{d\sim b} q^{\frac{L}{4}|d-c|} X_{L-1}(a|d,b)
\end{equation}
where the sum is over all $d$ such that $d\sim b$ admissible and $X_L$ 
is defined to be zero if $(b,c)$ do not satisfy both (\ref{weak}) and
(\ref{bound}).

The bosonic form for the configuration sum of the RSOS lattice models was 
first calculated by ABF \cite{ABF} for the case $N=1$ (in particular the 
Ising model and the hard square gas), by AB \cite{AB} for $N=2$ (for some
special value of $p$) and for general $N$ by the Kyoto group \cite{Date1}, 
\cite{Date2}.

Date et al. show in \cite{Date1}, \cite{Date2} that if one can find a solution
$f$ of the recursion relation
\begin{eqnarray}
\label{f_rec}
&&f_L(b,c)=\sum_{d} f_{L-1}(d,b)q^{L|d-c|/4}\\
\label{f_initial}
&&f_0(b,c)=\delta_{b,0}
\end{eqnarray}
where now the sum in (\ref{f_rec}) is over all $d$ such that the pair $(d,b)$ 
is weakly admissible then the solution to the problem (\ref{X_initial}), 
(\ref{X_rec}) is given by
\begin{equation}
\label{X_sol}
X_L(a|b,c)=q^{-\frac{a}{4}}(F_L(a|b,c)-F_L(-a|b,c))
\end{equation}
where
\begin{equation}
\label{def_F}
F_L(a|b,c)=\sum_{j\in {\bf Z}}q^{-pj^2+(\frac{p}{2}-a)j+\frac{a}{4}}
 f_L(b-a-2pj,c-a-2pj).
\end{equation}
Each term in the solution (\ref{X_sol}), (\ref{def_F}) of (\ref{X_initial}),
(\ref{X_rec}) can be interpreted by the sieving method \cite{Andrews2}. 
One starts with the solution $f_L$ for all 
weakly admissible paths. But one has overcounted since one is actually 
interested in paths obeying (\ref{bound}). Hence one subtracts off the 
generating function of paths which have at least one pair $(l_i,l_{i+1})$ with 
$l_i+l_{i+1}>2p-N-2$ and also the generating function of all path with at least
one pair with $l_i+l_{i+1}<N+2$. However, now one has subtracted too much
since there could be paths which have $l_i+l_{i+1}>2p-N-2$ \underline{and}
$l_j+l_{j+1}<N+2$ for some $i,j$. Hence one needs to add the generating 
function of paths which satisfy $l_i+l_{i+1}>2p-N-2$ and $l_j+l_{j+1}<N+2$
for $i<j$ and those which satisfy this condition for $i>j$ etc..

One may write (\ref{f_rec}) more explicitly as 
\begin{equation}
\label{f_recex}
f_L(b,c)=\sum_{k=0}^Nq^{\frac{L}{4}|N+b-c-2k|}f_{L-1}(b+N-2k,b).
\end{equation}
Hence a solution for $f_L$ is given by
\begin{equation}
\label{f_sol}
f_L^N(b,c)=q^{\frac{bc}{4N}}T_{\frac{N+b-c}{2}}^{(N)}(L,\frac{b}{2})
\end{equation}
since if we insert (\ref{f_sol}) into (\ref{f_recex}) we get
\begin{eqnarray}
q^{\frac{bc}{4N}}T_{\frac{N+b-c}{2}}^{(N)}(L,\frac{b}{2})&\!\!=&\!\!
 \sum_{k=0}^{N}q^{\frac{L}{4}|N+b-c-2k|}q^{\frac{b(b+N-2k)}{4N}}
 T_{\frac{2N-2k}{2}}(L-1,\frac{b+N-2k}{2})\nonumber\\
&\!\!=&\!\!q^{\frac{bc}{4N}}\sum_{k=0}^{N}q^{\frac{L}{4}|N+b-c-2k|+
 \frac{b(b-c+N-2k)}{4N}}T_{N-k}(L-1,\frac{N}{2}+\frac{b}{2}-k)
\end{eqnarray}
This exactly reduces to the recursion relation for the multinomials 
(\ref{multi_rec}) if we identify $n=\frac{N+b-c}{2}$ and $r=\frac{b}{2}$.
The initial conditions (\ref{f_initial}) are satisfied because of
(\ref{multi_initial}).

We would like to point out that (\ref{f_sol}) differs from the representation
of $f_L^N(b,b+N)$ in terms of multinomials given by Date et al. (cf. Theorem
4.2.5 in \cite{Date2}) which is not appropriate for taking the limit
$L\rightarrow\infty$.

Using (\ref{X_sol}), (\ref{def_F}) and (\ref{f_sol}) the full solution of the 
recursion equations (\ref{X_initial}), (\ref{X_rec}) is given by
\begin{eqnarray}
\label{X_multi}
X_L^N(a|b,c)=q^{\frac{(b-a)(c-a)}{4N}}\sum_j \left( 
 q^{\frac{j}{N}(jp(p-N)+p\frac{b+c-N}{2}-a(p-N))}
 T_{\frac{N+b-c}{2}}^{(N)}(L,\frac{b-a}{2}+pj)\right.\nonumber\\ \left.
 -q^{\frac{1}{N}(pj+a)(j(p-N)+\frac{b+c-N}{2})}T_{\frac{N+b-c}{2}}^{(N)}(L,
 \frac{b+a}{2}+pj)\right) .
\end{eqnarray}
Introducing
\begin{eqnarray}
\label{def_rsl}
&&r=\frac{1}{2}(b+c-N)\nonumber\\
&&s=a\nonumber\\
&&l=\frac{1}{2}(b-c+N)+1.
\end{eqnarray}
we can rewrite this expression in a more standard form
\begin{eqnarray}
\label{X_multi1}
\!\!\!\!\!\!\!\!\!\!\!\!\lefteqn{X_L^N(s|r+l-1,r-l+N+1)=}\nonumber\\
q^{\frac{(r-s+l-1)(r-s+N-l+1)}{4N}}&\sum_j& \left( q^{\frac{j}{N}(jpp'+pr-p's)}
 T_{l-1}^{(N)}(L,\frac{1}{2}(r-s+l-1)+pj)\right.\nonumber\\ 
 &-&\left. q^{\frac{1}{N}(pj+s)(jp'+r)}
 T_{l-1}^{(N)}(L,\frac{1}{2}(r+s+l-1)+pj)\right) .
\end{eqnarray}
where $p'=p-N$. This representation of the RSOS configuration sums in 
regime III generalizes the representations found by ABF \cite{ABF} and 
AB \cite{AB} for the cases $N=1$ and $N=2$, respectively.

According to Date et al. \cite{Date1}, \cite{Date2} $X_L(a|b,c)$ yields
the branching functions $c_{r,s}^{(l)}$ for the
$\widehat{su}(2)_{p-N-2}\times \widehat{su}(2)_{N}/\widehat{su}(2)_{p-2}$
coset models. In particular,
\begin{equation}
c_{r,s}^{(l)}=q^{-\eta}\lim_{L\rightarrow \infty, L\;{\rm even}}
 X_L(a|b,c)
\end{equation}
where $r$, $s$ and $l$ as defined in (\ref{def_rsl}). 
The exponent $\eta$ is given by
\begin{equation}
\label{def_eta}
\eta=\frac{1}{4}(b-a)-\gamma(r,l,s)
\end{equation}
where
\begin{equation}
\label{def_gamma}
\gamma(r,l,s)=\frac{r^2}{4(p-N)}+\frac{l^2}{4(N+2)}-\frac{1}{8}
 -\frac{s^2}{4p}.
\end{equation}
From (\ref{limit}) we see that $T^{(N)}_n(r)=\lim_{L\rightarrow\infty,
{\rm even}}T^{(N)}_n(L,r)$ only depends on $r\bmod N$. Further $T^{(N)}_n(r)$
has the symmetry $T^{(N)}_n(r)=T^{(N)}_n(n-r)$. Hence by defining
\begin{equation}
\label{def_m}
m_{r,s}(j)=|\frac{1}{2}(r-s)+pj \bmod N|
\end{equation}
where we use the convention that $y=x \bmod N$ lies in the intervall
$-N/2< y\leq N/2$ we may write
\begin{eqnarray}
\label{c_equ}
\lefteqn{
c_{r,s}^{(l)}=q^{-\eta}q^{\frac{1}{4N}(r+l-s-1)(r-l+N+1-s)}\sum_{m=0}^{N/2}
 T^{(N)}_{l-1}(m+\frac{l-1}{2})}\nonumber\\
&& \times \left\{\sum_{j\in {\bf Z},m_{r,s}(j)\equiv m} 
 q^{\frac{j}{N}(jpp'+pr-p's)}
 -\sum_{j\in {\bf Z},m_{r,-s}(j)\equiv m} q^{\frac{1}{N}(pj+s)(p'j+r)}\right\}
\end{eqnarray}
Notice that if $L$ even $b-a$ is always even. Hence we can conclude that 
$r-s$ is even (odd) if $l$ is odd (even). Therefore, the sum runs over $m$
integer if $l$ odd and halfinteger if $l$ even. This generalizes the
Neveu-Schwarz (NS) and Ramond (R) sectors that appear for $N=2$. We further 
see that $c_{r,s}^{(l)}$ factorizes into a parafermionic piece given by 
$T^{(N)}_{l-1}$ and a Rocha-Caridi type sum where the exponents give the 
positions of the singular vectors of the conformal field theory.

To make the connection with the branching functions given in \cite{KB} we may 
further introduce
\begin{equation}
\label{def_alpha}
\alpha_{r,s}(j)=\frac{(2pp'j+pr-p's)^2-N^2}{4Npp'}.
\end{equation}
In \cite{KB} the branching function $\chi_{r,s}^{(l)}$ is given by
\begin{equation}
\label{def_cKB}
\chi_{r,s}^{(l)}=\sum_{m=0}^{\frac{N}{2}}C_{2m}^l(q)\left\{
 \sum_{j\in{\bf Z},m_{r,s}(j)\equiv m}q^{\alpha_{r,s}(j)}
 - \sum_{j\in{\bf Z},m_{r,-s}(j)\equiv m}q^{\alpha_{r,-s}(j)}\right\}
\end{equation}
where $C_m^l$ are the $su(2)_N$ string functions \cite{KP}, \cite{JM}.
Here $m$ runs over integer if $l$ even (NS) and halfinteger if $l$ odd (R)
(which is exactly opposite to the convention used by Date et al. \cite{Date1}).

From (\ref{def_alpha}) follows that
\begin{eqnarray}
\label{alpha_equ}
\frac{j}{N}(jpp'+pr-p's)&=&\alpha_{r,s}(j)-\frac{1}{4Npp'}((pr-p's)^2-N^2)
\nonumber\\
\frac{1}{N}(pj+s)(p'j+r)&=&\alpha_{r,-s}(j)-\frac{1}{4Npp'}((pr-p's)^2-N^2).
\end{eqnarray}
Using (\ref{def_eta}), (\ref{def_gamma}) and (\ref{alpha_equ}) equation
(\ref{c_equ}) hence becomes
\begin{eqnarray}
\label{c_equ1}
\lefteqn{
c_{r,s}^{(l)}=q^{-\frac{1}{4N}+\frac{N}{4pp'}+\frac{l}{2N}-\frac{l^2}{2N(N+2)}
 -\frac{1}{8}}\sum_{m=0}^{N/2}T^{(N)}_{l-1}(m+\frac{l-1}{2})}\nonumber\\
 &&\times \left\{ \sum_{j\in{\bf Z}, m\equiv m_{r,s}(j)}q^{\alpha_{r,s}(j)}
 -\sum_{j\in{\bf Z}, m\equiv m_{r,-s}(j)}q^{\alpha_{r,-s}(j)}\right\}
\end{eqnarray}
We now want to make the connection between $T_{l-1}^{(N)}$ and the
$su(2)_N$ string functions $C_m^l$. According to Lepowsky and 
Primc \cite{LP} the relation between $T_{l-1}^{(N)}$ and the fermionic form
of the branching functions $b_m^l$ for the coset models 
$\widehat{su}(N)_1\times \widehat{su}(N)_1/\widehat{su}(N)_2$ is given as
\begin{eqnarray}
\lefteqn{q^{\frac{c}{24}}q^{-\frac{l(N-l)}{2N(N+2)}}b^l_{2Q-l}}\nonumber\\
&&=\sum_{m\in{\bf Z}^{N-1}, \frac{Q}{N}+e_1C^{-1}m\in{\bf Z}}
 q^{mC^{-1}m-e_lC^{-1}m}\prod_{i=1}^{N-1}\frac{1}{(q)_{m_i}}
\end{eqnarray}
where $c$ is the central charge $c=\frac{2(N-1)}{N+2}$.
In \cite{KP} and \cite{JM} one finds a representation of $b_m^l$ given
as a double sum also known as Hecke's indefinite modular form
\begin{eqnarray}
\label{def_b}
q^{\frac{c}{24}-h_m^l}b_m^l&=&\frac{1}{(q)_{\infty}^2}\left\{
 \left(\sum_{k\geq 0}\sum_{n\geq 0}-\sum_{k<0}\sum_{n<0} \right)
 (-)^kq^{\frac{k(k+1)}{2}+(l+1)n+\frac{(l+m)k}{2}+(N+2)(n+k)n}\right.
 \nonumber\\
&&+\left. \left(\sum_{k>0}\sum_{n\geq 0}-\sum_{k\leq 0}\sum_{n<0}\right)
 (-)^k q^{\frac{k(k+1)}{2}+(l+1)n+\frac{(l-m)k}{2}+(N+2)(n+k)n}\right\}
\end{eqnarray}
where $l=0,1,\ldots,N-1$, $l-m$ even and
\begin{equation}
\label{def_h}
h_m^l=\frac{l(l+2)}{4(N+2)}-\frac{m^2}{4N}
\end{equation}
The formula (\ref{def_b}) is only valid for $|m|\leq l$. For $|m|>l$ one may
use the symmetries
\begin{equation}
b_m^l=b_{-m}^l=b_{m+2N}^l=b_{N-m}^{N-l}
\end{equation}
This double sum (\ref{def_b}) is also related to the branching function
$e_{ml}^{(N)}$ of the coset models $\widehat{su}(2N)_1/\widehat{sp}(2N)_1$
\cite{Date1}, \cite{Date2}, \cite{KP}, \cite{JM} and in the notation of
\cite{Date1} we have $e_{ml}^{(N)}=b_m^l$. $e_{ml}^{(N)}$ in turn is related
to the string function of $\widehat{su}(2)_N$ via
\begin{equation}
e_{ml}^{(N)}=q^{\frac{-c}{24}}(q)_{\infty}C_m^l.
\end{equation}
Hence we find
\begin{eqnarray}
&&q^{-\frac{1}{4N}+\frac{N}{4pp'}+\frac{l}{2N}-\frac{l^2}{2N(N+2)}
 -\frac{1}{8}}T^{(N)}_{l-1}(m+\frac{l-1}{2})\nonumber\\
&&=\frac{1}{(q)_{\infty}}q^{-\frac{1}{4N}+\frac{N}{4pp'}+\frac{l}{2N}
 -\frac{l^2}{2N(N+2)}
 -\frac{1}{8}}q^{\frac{c}{24}}q^{-\frac{(l-1)(N-l+1)}{2N(N+2)}}b_{2m}^{l-1}
 \nonumber\\
&&=\frac{1}{(q)_{\infty}}q^{-\frac{1}{24}+\frac{N}{4pp'}}b_{2m}^{l-1}
 \nonumber\\
&&=\frac{1}{(q)_{\infty}}q^{-\frac{1}{24}+\frac{N}{4pp'}}e_{2m,l-1}^{(N)}
 \nonumber\\
&&=q^{\frac{-\hat{c}}{24}}C_{2m}^{l-1}.
\end{eqnarray}
where $\hat{c}=1-\frac{6N}{pp'}+\frac{2(N-1)}{N+2}$ is the central charge
of the coset $\widehat{su}(2)_{p-N-2}\times \widehat{su}(2)_N/
\widehat{su}(2)_{p-2}$. Hence we get for (\ref{c_equ1})
\begin{equation}
\label{c_equ2}
c_{r,s}^{(l)}=q^{\frac{-\hat{c}}{24}}\sum_{m=0}^{N/2}C_{2m}^{l-1}
 \left\{\sum_{j\in{\bf Z}, m\equiv m_{r,s}(j)}q^{\alpha_{r,s}(j)}
 -\sum_{j\in{\bf Z}, m\equiv m_{r,-s}(j)}q^{\alpha_{r,-s}(j)}\right\}
\end{equation}
and we see
\begin{equation}
c_{r,s}^{(l+1)}=q^{\frac{-\hat{c}}{24}}\chi_{r,s}^{(l)}.
\end{equation}
Therefore, the representation of the configuration sum $X_L(a|b,c)$ of the 
generalized RSOS models in terms of multinomials leads naturally to the 
branching functions of the coset models
$\widehat{su}(2)_{p-N-2}\times \widehat{su}(2)_N/\widehat{su}(2)_{p-2}$ 
\cite{KB} obtained by the Feigin and Fuchs construction \cite{FF} and
not to the representation given by Date et al. (see \cite{Date1} appendix A).

\section{Proof of the recursion relations for the multinomials}
\label{sec_proof}
\setcounter{equation}{0}

In this section we prove the recursion relations (\ref{multi_rec}) for 
the multinomials. First we notice that it is sufficient to prove 
(\ref{multi_rec}) for $n\leq \frac{N}{2}$ for all 
$r=-\frac{NL}{2},\ldots,\frac{NL}{2}$ since via the symmetry 
$T_n(L,-r)=T_{N-n}(L,r)$ one automatically gets (\ref{multi_rec}) for all 
$0\leq n \leq N$.

Let us now write out the right hand side of recursion relation 
(\ref{multi_rec}) in terms of the definition of the multinomials
\begin{eqnarray}
\sum_{k=0}^{n-1}q^{(n-k)(\frac{L}{2}+\frac{r}{N})}
 \sum_{\tilde{m}\in {\bf Z}_{\geq 0}^{N+1}}
 q^{mC^{-1}m-e_{N-k}C^{-1}m}(q)_{L-1}
 \prod_{i=0}^{N}\frac{1}{(q)_{m_i}}\nonumber\\
 +\sum_{k=n}^{N}q^{(k-n)(\frac{L}{2}-\frac{r}{N})}
 \sum_{\tilde{m}\in {\bf Z}_{\geq 0}^{N+1}}
 q^{mC^{-1}m-e_{N-k}C^{-1}m}(q)_{L-1}
 \prod_{i=0}^{N}\frac{1}{(q)_{m_i}}
\end{eqnarray}
where now the sums are restricted by
\begin{eqnarray}
\label{second_restra}
m_0=\frac{L-1}{2}-\frac{1}{2}-\frac{r-k}{N}-e_1C^{-1}m\in{\bf Z}_{\geq 0}\\
\label{second_restrb}
m_N=\frac{L-1}{2}+\frac{1}{2}+\frac{r-k}{N}-e_{N-1}C^{-1}m\in{\bf Z}_{\geq 0}
\end{eqnarray}
To transform the restrictions (\ref{second_restra}), (\ref{second_restrb})
into the restrictions (\ref{new_m0}), (\ref{new_mN}) one can make the 
variable change $m_k\rightarrow m_k-1$ in the $k^{\rm th}$ summand
for $k=1,2,\ldots,N-1$ ($k=0$ and $k=N$ remain unchanged) to obtain
\begin{eqnarray}
\label{rhs_rec}
\widetilde{\sum}_{\tilde{m}\in{\bf Z}_{\geq 0}^{N+1}}
 q^{mC^{-1}m}(q)_{L-1} \prod_{i=0}^{N}\frac{1}{(q)_{m_i}}
 &&\!\!\!\!\!\left\{ \sum_{k=0}^{n-1}
 q^{(n-k)(\frac{L}{2}+\frac{r}{N})-(2e_k+e_{N-k})C^{-1}m+g_{\frac{N}{2}}(k)}
 (1-q^{m_k})\right.\nonumber\\
 &&\!\!\!\!\!\left. +\sum_{k=n}^{N}q^{(k-n)(\frac{L}{2}
 -\frac{r}{N})-(2e_k+e_{N-k})C^{-1}m+g_{\frac{N}{2}}(k)}(1-q^{m_k})\right\}
 \nonumber\\
\end{eqnarray}
where 
\begin{equation}
\label{def_g}
g_{\frac{N}{2}}(k)=\left\{ \begin{array}{ll} k & {\rm for} \; 0\leq k\leq 
 \frac{N}{2} \\ N-k & {\rm for} \;\frac{N}{2}< k\leq N \end{array} \right. .
\end{equation}
This function arises from the variable change in the exponent when we use
\begin{equation}
e_iC^{-1}e_j=\frac{1}{N}\left\{ \begin{array}{ll}
 (N-i)j & {\rm for}\; j\leq i\\
 (N-j)i & {\rm for}\; j>i
\end{array} \right. .
\end{equation}

For the proof of (\ref{multi_rec}) we need to transform the expression 
in (\ref{rhs_rec}) step by step into the expression for $T_n^{(N)}(L,r)$ as 
given in (\ref{multi}). This is done by introducing four types of interpolating
functions $R_n^{(N)}(L,r,l)$ (for $l=n,n+1,\ldots,N$), $\bar{R}_n^{(N)}(L,r,l)$
(for $l=n-1,n-2,\ldots,0$), $G_n^{(N)}(L,r,l)$ (for $l=1,2,\ldots,n$) and 
$\bar{G}^{(N)}_n(L,r,l)$ (for $l=n-1,n,\ldots,N-1$). These interpolating 
functions have the following properties
\begin{eqnarray}
\label{prop_R}
R_n^{(N)}(L,r,l+1)=R_n^{(N)}(L,r,l)\\
\label{prop_Rb}
\bar{R}_n^{(N)}(L,r,l-1)=\bar{R}_n^{(N)}(L,r,l)
\end{eqnarray}
and $R_n^{(N)}(L,r,N)+\bar{R}_n^{(N)}(l,r,0)$ equals (\ref{rhs_rec}).
The properties of $G_n^{(N)}$ and $\bar{G}_n^{(N)}$ are as follows
\begin{eqnarray}
\label{prop_G}
G_n^{(N)}(L,r,l+1)=G_n^{(N)}(L,r,l)\\
\label{prop_Gb}
\bar{G}_n^{(N)}(L,r,l-1)=\bar{G}_n^{(N)}(L,r,l)
\end{eqnarray}
and $G_n^{(N)}(L,r,1)+\bar{G}_n^{(N)}(L,r,N-1)=T_n^{(N)}(L,r)$. And further
the relation between the $R$ and $G$ and the $\bar{R}$ and $\bar{G}$ 
interpolating functions are given as
\begin{eqnarray}
\label{RG_rel}
R_n^{(N)}(L,r,n)=G_n^{(N)}(L,r,n)\\
\label{RGb_rel}
\bar{R}_n^{(N)}(L,r,n-1)=\bar{G}_n^{(N)}(L,r,n-1).
\end{eqnarray}
The proof of recursion relation (\ref{multi_rec}) then follows from the string
of equations
\begin{eqnarray}
\label{string}
{\rm rhs\;of}\;(\ref{multi_rec})&=&
 R_n^{(N)}(L,r,N)+\bar{R}_n^{(N)}(L,r,0)\nonumber\\
&=&R_n^{(N)}(L,r,n)+\bar{R}_n^{(N)}(L,r,n-1)\;\;{\rm see}\; (\ref{prop_R}),
(\ref{prop_Rb})\nonumber\\
&=&G_n^{(N)}(L,r,n)+\bar{G}_n^{(N)}(L,r,n-1)\;\;{\rm see}\; (\ref{RG_rel}),
 (\ref{RGb_rel})\nonumber\\
&=&G_n^{(N)}(L,r,1)+\bar{G}_n^{(N)}(L,r,N-1)\;\;{\rm see}\; (\ref{prop_G}),
 (\ref{prop_Gb})\nonumber\\
&=&T_n^{(N)}(L,r).
\end{eqnarray}

Having outlined the strategy of the proof let us define
\begin{eqnarray}
\label{def_R}
\lefteqn{R^{(N)}_n(L,r,l)=\widetilde{\sum}_{\tilde{m}\in {\bf Z}^{N+1}}
 q^{mC^{-1}m}(q)_{L-1}\prod_{i=0}^{N}\frac{1}{(q)_{m_i}}}\nonumber\\
 && \times
 \left\{ \sum_{k=n}^{l-1}q^{(k-n)(\frac{L}{2}-\frac{r}{N})}A^{(N)}(L,r,l|k)
 +q^{(l-n)(\frac{L}{2}-\frac{r}{N})}\sum_{k=l}^{N}B^{(N)}(L,r,l|k)\right\}
\end{eqnarray}
for $l=n,n+1,\ldots,N$ where
\begin{eqnarray}
\label{def_A}
&&A^{(N)}(L,r,l|k)=q^{-(2e_k+e_{N-k})C^{-1}m+g_{\frac{N}{2}}(k)}(1-q^{m_k})\\
\label{def_B}
&&B^{(N)}(L,r,l|k)=\nonumber\\
&&\;\; q^{(-e_{N-l+1}+e_{k-l}+e_{k-l+1}-2e_k)C^{-1}m
 +g_{\frac{N}{2}}(l-1)-g_{[k-l+1,l-1]}(N-l+1)}(1-q^{m_k}).
\end{eqnarray}
and for $n\geq m$
\begin{equation}
\label{def_gg}
g_{m,n}(k)=\left\{ \begin{array}{ll} k & {\rm for} \; 0\leq k\leq m \\
 m & {\rm for} \; m<k<n \\
 m+n-k & {\rm for} \; n\leq k \leq n+m \\
 0 & {\rm otherwise} \end{array} \right. .
\end{equation}
We further define
\begin{equation}
g_{[m,n]}(k)=\left\{ \begin{array}{ll} g_{m,n}(k) & {\rm if}\; m\leq n\\
 g_{n,m}(k) & {\rm if}\; m>n \end{array} \right. .
\end{equation}
The sum $\widetilde{\sum}$ denotes from now on
always the sum over $\tilde{m}\in {\bf Z}^{N+1}$ such that (\ref{new_m0})
and (\ref{new_mN}) are satisfied and sums $\sum_{k=i}^j$ with $j<i$ are defined
to be zero. 

Similarly we define
\begin{eqnarray}
\label{def_Rbar}
\lefteqn{\bar{R}^{(N)}_n(L,r,l)=\widetilde{\sum}_{\tilde{m}\in {\bf Z}^{N+1}}
 q^{mC^{-1}m}(q)_{L-1}\prod_{i=0}^{N}\frac{1}{(q)_{m_i}}}\nonumber\\
 && \times \left\{
 \sum_{k=l+1}^{n-1}q^{(n-k)(\frac{L}{2}+\frac{r}{N})}\bar{A}^{(N)}(L,r,l|k)
 +q^{(n-l)(\frac{L}{2}+\frac{r}{N})}\sum_{k=0}^{l}\bar{B}^{(N)}(L,r,l|k)
 \right\}
\end{eqnarray}
where now $l=n-1,n-2,\ldots,0$. For $n=0$ $\bar{R}_n^{(N)}(L,r,l)$ is defined
to be zero. $\bar{A}$ and $\bar{B}$ are defined as
\begin{eqnarray}
\label{def_Abar}
&&\bar{A}^{(N)}(L,r,l|k)=q^{-(2e_k+e_{N-k})C^{-1}m+g_{\frac{N}{2}}(k)}
 (1-q^{m_k})\\
\label{def_Bbar}
&&\bar{B}^{(N)}(L,r,l|k)=\nonumber\\
&&\;\;q^{(-e_{N-l-1}+e_{N+k-l}+e_{N+k-l-1}-2e_k)C^{-1}m
 +g_{\frac{N}{2}}(l+1)-g_{[l-k+1,N-l-1]}(l+1)}(1-q^{m_k})
\end{eqnarray}

Notice that $R_n^{(N)}(L,r,N)+\bar{R}_n^{(N)}(L,r,0)$ equals (\ref{rhs_rec})
which was one of the conditions for $R$ and $\bar{R}$.
To prove (\ref{prop_R}) and (\ref{prop_Rb}) we need to make variable changes
in $m$. It will turn out to be useful to introduce
\begin{equation}
\label{def_E}
E_i=-e_{i-1}+2e_i-e_{i+1}
\end{equation}
for $i=0,1,2,\ldots,N$ (remember that $e_i=0$ if 
$i\not\in \{1,2,\ldots,N-1\}$ ).
Since $E_i$ is the $i^{\rm th}$ column of $C$ for $i=1,2,\ldots,N-1$ we have
\begin{equation}
\label{prop_E}
C^{-1}E_i=e_i, \; {\rm for}\; i=1,2,\ldots,N-1.
\end{equation}
Let us further define for $m\leq l$
\begin{equation}
\label{def_Etilde}
\tilde{E}_{m,l}\equiv -\sum_{i=m}^{l}E_i=e_{m-1}-e_m-e_l+e_{l+1}.
\end{equation}
One may derive for $i\leq j$
\begin{eqnarray}
\label{prop_Etilde}
\sum_{k=i-l+1}^i\tilde{E}_{k,i+j-k}&=&e_{i-l}-e_i-e_j+e_{j+l}\nonumber\\
 &=&-\left( \sum_{k=i-l}^{i}(k-i+l)E_k+l\sum_{k=i+1}^{j-1}E_k
 +\sum_{k=j}^{j+l}(j+l-k)E_k \right) .
\end{eqnarray}
To prove (\ref{prop_R}) we rewrite all terms $B^{(N)}(L,r,l+1|k)$ 
($k=l+1,\dots,N$) in $R_n^{(N)}(L,r,l+1)$ as follows
\begin{eqnarray}
\label{trick}
\lefteqn{q^{\frac{L}{2}-\frac{r}{N}}B^{(N)}(L,r,l+1|k)}\nonumber\\
&&=q^{\frac{L}{2}-\frac{r}{N}-m_0}B^{(N)}(L,r,l+1|k)
 -q^{\frac{L}{2}-\frac{r}{N}-m_0}B^{(N)}(L,r,l+1|k)(1-q^{m_0})\nonumber\\
&&=q^{(e_1-e_{N-l}+e_{k-l-1}+e_{k-l}-2e_k)C^{-1}m+g_{\frac{N}{2}}(l)
 -g_{[k-l,l]}(N-l)}(1-q^{m_k})\nonumber\\
&&-q^{(e_1-e_{N-l}+e_{k-l-1}+e_{k-l}-2e_k)C^{-1}m+g_{\frac{N}{2}}(l)
 -g_{[k-l,l]}(N-l)}(1-q^{m_k})(1-q^{m_0})
\end{eqnarray}
remembering that $m_0$ is given by (\ref{new_m0}). In the second term one can
make the variable change $m\rightarrow m+v$ with
\begin{eqnarray}
\label{var_change}
v&=&-e_{k-l}-e_l+e_k=\sum_{j=1}^{{\rm min}(k-l,l)}\tilde{E}_{j,k-j}\nonumber\\
&=&\left\{ \begin{array}{ll}
-\left(\sum_{j=1}^{k-l}jE_j+(k-l)\sum_{j=k-l+1}^{l-1}E_j
 +\sum_{j=l}^{k-1}(k-j)E_j\right)
& {\rm for}\; k-l\leq l \\
-\left( \sum_{j=1}^{l}jE_j+l\sum_{j=l+1}^{k-l-1}E_j
 +\sum_{j=k-l}^{k-1}(k-j)E_j\right)
& {\rm for}\; k-l> l \end{array}\right.
\end{eqnarray}
where we used (\ref{prop_Etilde}). Hence under the
sum $\widetilde{\sum}_{\tilde{m}\in {\bf Z}^{N+1}}q^{mC^{-1}m}(q)_{L-1}
\prod_{i=0}^N\frac{1}{(q)_{m_i}}$ (in the following we abbreviate this
expression by ``under the sum $\widetilde{\sum}$'') the second term 
in (\ref{trick}) becomes
\begin{eqnarray}
-q^{(e_1-e_{N-l}-2e_l+e_{k-l-1}-e_{k-l})C^{-1}m+g_{\frac{N}{2}}(l)}
(1-q^{m_l})
\left\{ \begin{array}{rl}
 (1-q^{m_{k-l}}) & {\rm for}\; k-l<l\\
 (1-q^{m_{k-l}-1}) & {\rm for}\; k-l=l\\
q^{-1}(1-q^{m_{k-l}}) & {\rm for}\; k-l>l
\end{array} \right.
\end{eqnarray}
We see that in the sum $\sum_{k=l+1}^N$ these terms cancel pairwise
(the $-q^{m_{k-l}}$ in the $k^{\rm th}$ term cancels the $1$ in the
$(k+1)^{\rm th}$ term since $m_i=E_iC^{-1}m=(-e_{i-1}+2e_i-e_{i+1})C^{-1}m$).
Hence only the $1$ for $k=l+1$ and $-q^{m_{N-l}}$ for $k=N$ remain.
Altogether we get
\begin{eqnarray}
\!\!\!\lefteqn{R_n^{(N)}(L,r,l+1)=\widetilde{\sum}_{\tilde{m}\in {\bf Z}^{N+1}}
 q^{mC^{-1}m}(q)_{L-1}\prod_{i=0}^N\frac{1}{(q)_{m_i}}}\nonumber\\
&&\times \left\{ \sum_{k=n}^l q^{(k-n)(\frac{L}{2}-\frac{r}{N})}
 A^{(N)}(L,r,l+1|k)
 \right. \nonumber\\
&&+q^{(l+1-n)(\frac{L}{2}-\frac{r}{N})}\sum_{k=l+1}^N q^{-m_0} 
 B^{(N)}(L,r,l+1|k)\nonumber\\
&&-q^{(l-n)(\frac{L}{2}-\frac{r}{N})+(-e_{N-l}-2e_l)C^{-1}m+g_{\frac{N}{2}}(l)}
 (1-q^{m_l})\nonumber\\
&&\left. +q^{(l-n)(\frac{L}{2}-\frac{r}{N})+(e_1-e_{N-l+1}-2e_l)C^{-1}m
 +g_{\frac{N}{2}}(l)-\theta(N-2l)}(1-q^{m_l})\right\}
\end{eqnarray}
where $\theta(x)=1$ for $x\geq 0$ and $0$ otherwise. But
\begin{equation}
A^{(N)}(L,r,l+1|l)=q^{-(2e_l+e_{N-l})C^{-1}m+g_{\frac{N}{2}}(l)}(1-q^{m_l})
\end{equation}
and
\begin{equation}
B^{(N)}(L,r,l|l)=q^{(-e_{N-l+1}+e_1-2e_l)C^{-1}m+g_{\frac{N}{2}}(l-1)
 -g_{[1,l-1]}(N-l+1)}(1-q^{m_l})
\end{equation}
and $A^{(N)}(L,r,l+1|k)=A^{(N)}(L,r,l|k)$ for $k<l$. One may further
show that
\begin{equation}
g_{\frac{N}{2}}(l)-\theta(N-2l)=g_{\frac{N}{2}}(l-1)-g_{1,l-1}(N-l+1).
\end{equation}
Hence we have proven that
\begin{eqnarray}
\label{R_equh}
\!\!\!\lefteqn{R_n^{(N)}(L,r,l+1)=\widetilde{\sum}_{\tilde{m}\in {\bf Z}^{N+1}}
 q^{mC^{-1}m}(q)_{L-1}\prod_{i=0}^N\frac{1}{(q)_{m_i}}}\nonumber\\
&\times &
 \left\{ \sum_{k=n}^{l-1} q^{(k-n)(\frac{L}{2}-\frac{r}{N})}A^{(N)}(L,r,l|k)
 \right. \nonumber\\
&&+q^{(l+1-n)(\frac{L}{2}-\frac{r}{N})}\sum_{k=l+1}^N q^{-m_0} 
 B^{(N)}(L,r,l+1|k)\nonumber\\
&&\left. +q^{(l-n)(\frac{L}{2}-\frac{r}{N})}B^{(N)}(L,r,l|l) \right\}
\end{eqnarray}

To prove (\ref{prop_R}) it therefore remains to show that under the sum 
$\widetilde{\sum}$
\begin{equation}
q^{(\frac{L}{2}-\frac{r}{N})}\sum_{k=l+1}^N q^{-m_0}B^{(N)}(L,r,l+1|k)
=\sum_{k=l+1}^N B^{(N)}(L,r,l|k).
\end{equation}
To this end we define the following function $H$ for $d\leq N-l$ and 
$p=d,d+1,\ldots,N-l$
\begin{equation}
\label{def_H}
H^{(N)}(L,r,l|d,p)=\sum_{k=l+p}^N q^{-\sum_{i=d}^{p-1}m_i-e_{d-1}C^{-1}m}
 B^{(N)}(L,r,l+1|k).
\end{equation}
$H$ has the property that under the sum $\widetilde{\sum}$
\begin{eqnarray}
\label{H_equ}
H^{(N)}(L,r,l|d,p)&=&H^{(N)}(L,r,l|d,p+1)+q^{-e_d C^{-1}m}B^{(N)}(L,r,l|p+l)
\nonumber\\
&=&H^{(N)}(L,r,l|d,p+1)\nonumber\\
&+&q^{(-e_{N-l+1}-e_d+e_p+e_{p+1}-2e_{p+l})C^{-1}m+g_{\frac{N}{2}}(l-1)
 -g_{[p+1,l-1]}(N-l+1)}(1-q^{m_{p+l}})\nonumber\\
\end{eqnarray}
The proof of this formula is similar to the derivation of (\ref{R_equh}).
One rewrites all the terms in the sum and makes appropriate variable changes.
For the details of the proof we refer the reader to appendix \ref{app_proofH}.

Using (\ref{H_equ}) we have under the sum $\widetilde{\sum}$
\begin{eqnarray}
&&q^{(\frac{L}{2}-\frac{r}{N})}\sum_{k=l+1}^N q^{-m_0}B^{(N)}(L,r,l+1|k)
 \nonumber\\
&=&q^{(\frac{L}{2}-\frac{r}{N})-m_0}H^{(N)}(L,r,l|1,1)\nonumber\\
&=&q^{(\frac{L}{2}-\frac{r}{N})-m_0}\left\{ H^{(N)}(L,r,l|1,N-l)
 +\sum_{p=1}^{N-l-1}q^{-e_1C^{-1}m}B^{(N)}(L,r,l|p+l)\right\}\nonumber\\
&=&q^{e_1C^{-1}m}\left\{q^{-\sum_{i=1}^{N-l-1}m_i}q^{e_{N-l-1}C^{-1}m
 +g_{\frac{N}{2}}(l)-g_{[N-l,l]}(N-l)}(1-q^{m_N})\right.\nonumber\\
&&\left. +\sum_{k=l+1}^{N-1}q^{-e_1C^{-1}m}B^{(N)}(L,r,l|k)\right\}\nonumber\\
&=&\sum_{k=l+1}^NB^{(N)}(L,r,l|k).
\end{eqnarray}
For the last equality we used $g_{\frac{N}{2}}(l)-g_{[N-l,l]}(N-l)=0$ and 
$-\sum_{i=1}^{N-l-1}m_i=(-e_1-e_{N-l-1}+e_{N-l})C^{-1}m$ which follows from
(\ref{def_Etilde}). Hence we conclude
\begin{equation}
R_n^{(N)}(L,r,l+1)=R_n^{(N)}(L,r,l).
\end{equation}

The proof of (\ref{prop_Rb}) is analogous (one can actually obtain
(\ref{prop_Rb}) from (\ref{prop_R}) by dropping the term $k=n$ and then
replacing $r\rightarrow -r$, $m_i \rightarrow m_{N-i}$ and identifying 
$n_{\rm new}=N-n$ and $l_{\rm new}=N-l$).

Notice that for $n=0$ the proof of (\ref{multi_rec}) is already complete
since
\begin{eqnarray}
\lefteqn{{\rm rhs\; of}\;(\ref{multi_rec})}\nonumber\\
&&=R_0^{(N)}(L,r,N)+\bar{R}_0^{(N)}(L,r,0) \nonumber\\
&&=R_0^{(N)}(L,r,N)=R_0^{(N)}(L,r,1) \;\;{\rm via}\;(\ref{prop_R})\nonumber\\
&&=\widetilde{\sum_{\tilde{m}\in {\bf Z}^{N+1}}}q^{mC^{-1}m}(q)_{L-1}
 \prod_{i=0}^N \frac{1}{(q)_{m_i}} \underbrace{ \left\{
 (1-q^{m_0})+q^{\frac{L}{2}-\frac{r}{N}}
 \sum_{k=1}^N q^{(e_{k-1}-e_k)C^{-1}m}(1-q^{m_k}) \right\}}_{1-q^L}
\nonumber\\
\end{eqnarray}
Again all terms cancel pairwise except the $1$ in the first term and the
$-q^{m_N}$ in the last term ($k=N$). Using (\ref{new_mN}) one obtains
$1-q^L$ for the terms in the bracket which can be combined with
$(q)_{L-1}$ to $(q)_L$. Hence
\begin{equation}
{\rm rhs\;of}\;(\ref{multi_rec})=
\widetilde{\sum}_{\tilde{m}\in {\bf Z}^{N+1}}q^{mC^{-1}m}(q)_{L}
 \prod_{i=0}^N \frac{1}{(q)_{m_i}}
=T_0^{(N)}(L,r)
\end{equation}
and (\ref{multi_rec}) is proven for $n=0$.

For $n>0$ we need to introduce the functions $G_n^{(N)}(L,r)$ and 
$\bar{G}_n^{(N)}(L,r)$. Let us define
\begin{equation}
\label{def_G}
G_n^{(N)}(L,r,l)=
\widetilde{\sum}_{\tilde{m}\in {\bf Z}^{N+1}}q^{mC^{-1}m}(q)_{L-1}
 \prod_{i=0}^N \frac{1}{(q)_{m_i}} \left\{ \sum_{k=n}^N q^{-e_{n-l}C^{-1}m}
B^{(N)}(L,r,l|k)\right\}
\end{equation}
for $l=1,2,\ldots,n$ and similarly
\begin{equation}
\label{def_Gbar}
\bar{G}^{(N)}_n(L,r,l)=q^{\frac{L}{2}+\frac{r}{N}}
 \widetilde{\sum_{\tilde{m}\in {\bf Z}^{N+1}}}
 q^{mC^{-1}m}(q)_{L-1}\prod_{i=0}^N \frac{1}{(q)_{m_i}} 
 \left\{ \sum_{k=0}^{n-1} q^{-e_{N+n-l-1}C^{-1}m}\bar{B}^{(N)}(L,r,l|k) 
\right\}
\end{equation}
for $l=n-1,n,\ldots,N-1$. Notice that with these definitions 
equations (\ref{RG_rel}) and (\ref{RGb_rel}) hold. 

To show (\ref{prop_G}) we may write
\begin{eqnarray}
\lefteqn{G_n^{(N)}(L,r,l+1)}\nonumber\\
&&=\widetilde{\sum}_{\tilde{m}\in {\bf Z}^{N+1}}
 q^{mC^{-1}m}(q)_{L-1}\prod_{i=0}^N \frac{1}{(q)_{m_i}} 
 \left\{ \sum_{k=n}^N q^{-e_{n-l-1}C^{-1}m} B^{(N)}(L,r,l+1|k)\right\}
 \nonumber\\
&&=\widetilde{\sum}_{\tilde{m}\in {\bf Z}^{N+1}}
 q^{mC^{-1}m}(q)_{L-1}\prod_{i=0}^N \frac{1}{(q)_{m_i}} 
 H^{(N)}(L,r,l|n-l,n-l).
\end{eqnarray}
Via (\ref{H_equ}) we get under the sum $\widetilde{\sum}$
\begin{eqnarray}
\lefteqn{H^{(N)}(L,r,l|n-l,n-l)}\nonumber\\
&&=H^{(N)}(L,r,l|n-l,N-l)
 +\sum_{p=n-l}^{N-l-1}q^{-e_{n-l}C^{-1}m}B^{(N)}(L,r,l|p+l)\nonumber\\
&&=q^{-\sum_{i=n-l}^{N-l-1}m_i-e_{n-l-1}C^{-1}m}B^{(N)}(L,r,l+1|N)
 +\sum_{k=n}^{N-1}q^{-e_{n-l}C^{-1}m}B^{(N)}(L,r,l|k)\nonumber\\
\end{eqnarray}
Using now $-\sum_{i=n-l}^{N-l-1}m_i=
(e_{n-l-1}-e_{n-l}-e_{N-l-1}+e_{N-l})C^{-1}m$ via (\ref{def_Etilde}) we see
that
\begin{eqnarray}
G_n^{(N)}(L,r,l+1)&=&
\widetilde{\sum}_{\tilde{m}\in {\bf Z}^{N+1}}
 q^{mC^{-1}m}(q)_{L-1}\prod_{i=0}^N \frac{1}{(q)_{m_i}} 
\sum_{k=n}^{N}q^{-e_{n-l}C^{-1}m}B^{(N)}(L,r,l|k)\nonumber\\
&=&G_n^{(N)}(L,r,l).
\end{eqnarray}
The derivation of (\ref{prop_Gb}) is similar to the one for (\ref{prop_G}).

Finally, we need to show that $G_n^{(N)}(L,r,1)+\bar{G}_n^{(N)}(L,r,N-1)
=T_n^{(N)}(L,r)$. To this end we may write
\begin{eqnarray}
\label{G1_equ}
\lefteqn{G_n^{(N)}(L,r,1)}\nonumber\\
&=&\widetilde{\sum}_{\tilde{m}\in {\bf Z}^{N+1}}
 q^{mC^{-1}m}(q)_{L-1}\prod_{i=0}^N \frac{1}{(q)_{m_i}} 
 \left\{ \sum_{k=n}^N q^{-e_{n-1}C^{-1}m}B^{(N)}(L,r,1|k)\right\}
 \nonumber\\
&=&\widetilde{\sum}_{\tilde{m}\in {\bf Z}^{N+1}}
 q^{mC^{-1}m}(q)_{L-1}\prod_{i=0}^N \frac{1}{(q)_{m_i}} 
 \sum_{k=n}^N q^{(-e_{n-1}+e_{k-1}-e_k)C^{-1}m}(1-q^{m_k})\nonumber\\
&=&\widetilde{\sum}_{\tilde{m}\in {\bf Z}^{N+1}}
 q^{mC^{-1}m}(q)_{L-1}\prod_{i=0}^N \frac{1}{(q)_{m_i}} 
 \left\{ q^{-e_n C^{-1}m}-q^{(\frac{L}{2}+\frac{r}{N})-e_{n-1}C^{-1}m}
 \right\}
\end{eqnarray}
since the terms in the sum $\sum_{k=n}^N$ again cancel pairwise except for
the most outer terms. Similarly
\begin{equation}
\label{GbarN_equ}
\bar{G}_n^{(N)}(L,r,N-1)=q^{\frac{L}{2}+\frac{r}{N}}
 \widetilde{\sum_{\tilde{m}\in {\bf Z}^{N+1}}}
 q^{mC^{-1}m}(q)_{L-1}\prod_{i=0}^N \frac{1}{(q)_{m_i}} 
 \left\{ q^{-e_{n-1}C^{-1}m}-q^{\frac{L}{2}-\frac{r}{N}-e_n C^{-1}m}\right\}
\end{equation}
Inserting (\ref{G1_equ}) and (\ref{GbarN_equ}) into (\ref{string}) yields
\begin{eqnarray}
\label{superfinal}
{\rm rhs\;of}\;(\ref{multi_rec})&=&
 \widetilde{\sum}_{\tilde{m}\in {\bf Z}^{N+1}}
 q^{mC^{-1}m}(q)_{L-1}\prod_{i=0}^N \frac{1}{(q)_{m_i}} 
 \left\{ q^{-e_n C^{-1}m}-q^{L-e_n C^{-1}m} \right\}\nonumber\\
&&=T_n^{(N)}(L,r)
\end{eqnarray}
and hence (\ref{multi_rec}) is proven.

\section{Summary and Discussion}
\label{summary}
\setcounter{equation}{0}

We have given explicit formulas for the multinomials in (\ref{multi})
which generalize the trinomials first introduced by Andrews and 
Baxter \cite{AB}. These multinomials satisfy depth one recursion 
relations (\ref{multi_rec}) (depth one since (\ref{multi_rec}) relates $T$ at
$L$ to $T$'s at $L-1$). These recursion relations have been proven in 
section \ref{sec_proof}. Using the recursion relations we were able to 
express the configuration sums for the general RSOS models previously studied 
by Date et al. \cite{Kyoto}-\cite{Date2} in terms of multinomials (see 
(\ref{X_multi1})). These formulas differ from the configuration sums obtained
by Date et al.. Whereas Date et al. \cite{Date1}, \cite{Date2} get
double sum expressions for the branching functions of the
$\widehat{su}(2)_{p-N-2}\times \widehat{su}(2)_{N}/\widehat{su}(2)_{p-2}$ coset
(where $p$ and $N$ were introduced in section \ref{sec_partition}) which
in turn can be expressed in terms of elliptic theta functions,
the expressions in terms of the multinomials yield formulas of the 
Rocha-Caridi type. They factor into a parafermionic piece and a Rocha-Caridi 
piece (see (\ref{c_equ1})). With the help of relations between string functions
and branching functions of certain models we were able to express
the parafermionic piece in terms of the string functions of 
$\widehat{su}(2)_N$. The so obtained branching functions for the
$\widehat{su}(2)_{p-N-2}\times \widehat{su}(2)_{N}/\widehat{su}(2)_{p-2}$ coset
models are of the form given in \cite{KB}.

Even though it sufficed for our purposes to prove a depth one recursion
relation for the multinomials there are other recursion
relations of higher depth. Andrews and Baxter \cite{AB} found that 
the trinomials satisfy depth one recursion relations which mix the different 
types $T_1$ and $T_0$. The recursion relations (\ref{multi_rec}) also mix
the different types of multinomials $T^{(N)}_n$ with $n=0,1,\ldots,N-1$.
But in addition there exist recursion relations for the trinomials which
become tautologies for $q=1$ and recursion relations of higher depths 
some of which only involve the same type of trinomials \cite{AB}, 
\cite{Andrews}, \cite{Andrews1}, \cite{BMO}.
The recursion relations for the trinomials of higher depth were used in
\cite{BMO} to get the bosonic form for the characters $SM(2,4\nu)$.

As already mentioned in the introduction one might hope to obtain polynomial 
expressions for the coset constructions 
$\widehat{su}(2)_N\times \widehat{su}(2)_M/\widehat{su}(2)_{N+M}$
with fractional level $M$ in terms of multinomials.
Date et al. \cite{Date3} give an expression of the configuration sum of higher
rank coset constructions in terms of multinomials.
One might further speculate that multinomials with the matrix $C^{-1}$
replaced by a different matrix $B$ lead to useful generalizations of the 
multinomials introduced in this paper.

Further it is interesting to explore the connections and implications of the 
multinomials for partition theory and hypergeometric functions.
A partition interpretation in terms of Durfee dissections of a partition
has been given in \cite{Ole}.
There is a vast number of partition interpretations for Rogers-Ramanujan
type identities (e.g. \cite{Schur}, \cite{Andrews2}, \cite{ABBBFV},
\cite{Burge}, \cite{Burge1}, \cite{FW} and references therein).
In \cite{ABBBFV} Andrews et al. find expressions for the generating functions
of partitions with prescribed hook differences which yield the characters
for the minimal models $M(p,p')$ when $p$ and $p'$ are coprime.
A partition theoretic interpretation for the 
$\widehat{su}(2)_{M}\times \widehat{su}(2)_{N}/\widehat{su}(2)_{M+N}$ coset
models should also exist.

\section*{Acknowledgements}

I would like to thank Barry McCoy for many helpful discussions and comments.
Further I like to thank Ole Warnaar for private communications and Alexander
Berkovich for helpful comments on the manuscript and discussions.

This work was partially supported by NSF grant DMR-9404747.

\appendix

\section{Appendix}
\label{app_proofH}
\setcounter{equation}{0}

In this appendix we want to prove equation (\ref{H_equ}), namely that under
the sum $\widetilde{\sum}$
\begin{eqnarray}
H^{(N)}(L,r,l|d,p)&=&H^{(N)}(L,r,l|d,p+1)+q^{-e_d C^{-1}m}B^{(N)}(L,r,l|p+l)
\nonumber\\
&=&H^{(N)}(L,r,l|d,p+1)\nonumber\\
&+&q^{(-e_{N-l+1}-e_d+e_p+e_{p+1}-2e_{p+l})C^{-1}m+g_{\frac{N}{2}}(l-1)
 -g_{[p+1,l-1]}(N-l+1)}(1-q^{m_{p+l}})\nonumber\\
\end{eqnarray}
We start with the left hand side of this equation and write out 
explicitly the definition of $H$ according to (\ref{def_H})
\begin{eqnarray}
\label{H_ex}
\lefteqn{H^{(N)}(L,r,l|d,p)}\nonumber\\
&=& \sum_{k=l+p}^N q^{(-e_d-e_{p-1}+e_p-e_{N-l}+e_{k-l-1}+e_{k-l}-2e_k)C^{-1}m
 +g_{\frac{N}{2}}(l)-g_{[k-l,l]}(N-l)}(1-q^{m_k}).
\end{eqnarray}
Let us take all terms in the sum with $k=l+p+1,l+p+2,\ldots,N$ and rewrite them
as
\begin{eqnarray}
\label{trick1}
&&q^{(-e_d-e_{p-1}+e_p-e_{N-l}+e_{k-l-1}
 +e_{k-l}-2e_k)C^{-1}m+g_{\frac{N}{2}}(l)-g_{[k-l,l]}(N-l)}(1-q^{m_k})
 \nonumber\\
&=&q^{(-e_d-e_{p}+e_{p+1}-e_{N-l}+e_{k-l-1}
 +e_{k-l}-2e_k)C^{-1}m+g_{\frac{N}{2}}(l)-g_{[k-l,l]}(N-l)}(1-q^{m_k})
 \nonumber\\
&-&q^{(-e_d-e_{p}+e_{p+1}-e_{N-l}+e_{k-l-1}
 +e_{k-l}-2e_k)C^{-1}m+g_{\frac{N}{2}}(l)-g_{[k-l,l]}(N-l)}(1-q^{m_k})
 (1-q^{m_p}) \nonumber\\&&
\end{eqnarray}
Now we make the variable change $m\rightarrow m+v$ in the second term 
of (\ref{trick1}) where
\begin{eqnarray}
\label{var_change1}
v&\!\!\!=&\!\!\!e_p-e_{k-l}-e_{l+p}+e_k\nonumber\\
&\!\!\!=&\!\!\!\left\{ \begin{array}{ll} 
 \sum_{j=p+1}^{k-l}\tilde{E}_{j,k+p-j} & {\rm for}\;k-l<l+p\\
 \sum_{j=p+1}^{l+p}\tilde{E}_{j,k+p-j} & {\rm for}\;k-l\geq l+p
\end{array} \right. \nonumber\\
&\!\!\!=&\!\!\!\!\left\{ \!\!\!\!\begin{array}{ll} 
 -\left( \sum_{j=p}^{k-l}(j-p)E_j+(k-l-p)\sum_{j=k-l+1}^{l+p-1}E_j
 +\sum_{j=l+p}^k (k-j)E_j \right) & \!{\rm if}\;k-l<l+p\\
 -\left( \sum_{j=p}^{l+p} (j-p)E_j+l\sum_{j=l+p+1}^{k-l-1}E_j
 +\sum_{j=k-l}^k (k-j)E_j \right) & \!{\rm if}\;k-l>l+p\\
 -\left( \sum_{j=p}^{l+p} (j-p)E_j+\sum_{j=l+p+1}^k (k-j)E_j \right) 
 & \!{\rm if}\;k-l=l+p
\end{array} \right.\nonumber\\ && 
\end{eqnarray}
where we used (\ref{prop_Etilde}) and the fact that $k=l+p+1,\ldots,N$ and
hence $k-l>p$. Therefore, after making the variable change (\ref{var_change1})
the second term of (\ref{trick1}) becomes under the sum $\widetilde{\sum}$
\begin{eqnarray}
\label{after_varchange}
-q^{(e_p+e_{p+1}-e_d-e_{N-l}-2e_{p+l}+e_{k-l-1}-e_{k-l})C^{-1}m
+g_{\frac{N}{2}}(l)-g_{[k-l,l]}(N-l)}(1-q^{m_{l+p}})\nonumber\\
\times \left\{ \begin{array}{ll}
 q^{g_{k-l-p,l}(N-l-p)}(1-q^{m_{k-l}}) & {\rm for}\;k-l<l+p\\
 q^{g_{l,l}(N-l-p)}(1-q^{m_{k-l}-1}) & {\rm for}\;k-l=l+p\\ 
 q^{g_{l,k-l-p}(N-l-p)-1}(1-q^{m_{k-l}}) & {\rm for}\;k-l>l+p
\end{array} \right. .
\end{eqnarray}
Notice that
\begin{eqnarray}
\label{g_equ}
g_{[k-l-p,l]}(N-l-p)-g_{[k-l,l]}(N-l)&=& \left\{ \begin{array}{ll}
 -p & {\rm for} \; N-l\leq l\\
 N-p-2l & {\rm for} \; l<N-l<l+p\\
 0 & {\rm for} \; l+p\leq N-l
\end{array} \right.\nonumber\\
&=&-g_{[p,l]}(N-l)
\end{eqnarray}
where we used that $k=l+p+1,\ldots,N$ and hence $N-l-p\geq k-l-p$.
From (\ref{g_equ}) we see that $g_{[k-l-p,l]}(N-l-p)-g_{[k-l,l]}(N-l)$ 
is $k$ independent. Therefore it follows again that under the 
sum $\sum_{k=l+p+1}^N$ all terms in (\ref{after_varchange}) cancel pairwise 
except for the $1$ for $k=l+p+1$ and the $-q^{m_{k-l}}$ for $k=N$. The sum
$\sum_{k=l+p+1}^N$ of the first term in (\ref{trick1}) gives exactly
$H^{(N)}(L,r,l|d,p+1)$. Hence
\begin{eqnarray}
\label{H_der}
\lefteqn{H^{(N)}(L,r,l|d,p)}\nonumber\\
&=&H^{(N)}(L,r,l|d,p+1)+(1-q^{m_{p+l}}) \{\nonumber\\
&&+q^{(2e_p-e_d-e_{N-l}-2e_{l+p})C^{-1}m+g_{\frac{N}{2}}(l)-g_{[p,l]}(N-l)}
 \nonumber\\
&&-q^{(2e_p-e_d-e_{N-l}-2e_{p+l})C^{-1}m+g_{\frac{N}{2}}(l)+g_{[p+1,l]}(N-l-p)
 -g_{1,l}(N-l)}\nonumber\\
&&+q^{(e_p+e_{p+1}-e_d-2e_{p+l}-e_{N-l+1})C^{-1}m+g_{\frac{N}{2}}(l)
 -g_{[N-l,l]}(N-l)+g_{[l,N-l-p]}(N-l-p)-\theta(N-2l-p)} \}\nonumber\\
\end{eqnarray}
where the second term comes from the term $k=l+p$ in (\ref{H_ex}).
From (\ref{g_equ}) we can immediately see that the second and third term
in (\ref{H_der}) cancel. To prove (\ref{H_equ}) it merely remains to show
that
\begin{eqnarray}
&&g_{\frac{N}{2}}(l)-g_{[N-l,l]}(N-l)+g_{[l,N-l-p]}(N-l-p)-\theta(N-2l-p)
\nonumber\\
&=&g_{\frac{N}{2}}(l-1)-g_{[p+1,l-1]}(N-l+1).
\end{eqnarray}
But $g_{\frac{N}{2}}(l)=g_{[N-l,l]}(N-l)$ and
\begin{eqnarray}
\lefteqn{g_{[l,N-l-p]}(N-l-p)-\theta(N-2l-p)}\nonumber\\
&=&\left\{ \begin{array}{ll}
 N-l-p & {\rm for}\; N-l-p\leq l \\
 l & {\rm for}\; N-l-p\geq l
\end{array} \right.
-\left\{ \begin{array}{ll}
 1 & {\rm for}\; N-l-p\geq l \\
 0 & {\rm for}\; N-l-p<l
\end{array} \right. \nonumber\\
&=&\left\{ \begin{array}{ll}
 N-l-p & {\rm for}\; N-l-p<l \\
 l-1 & {\rm for}\; N-l-p\geq l
\end{array} \right.
\end{eqnarray}
and
\begin{eqnarray}
\lefteqn{g_{\frac{N}{2}}(l-1)-g_{[p+1,l-1]}(N-l+1)}\nonumber\\
&=&\left\{ \begin{array}{ll}
 l-1 & {\rm for}\; l-2\leq N-l \\
 N-l+1 & {\rm for}\; N-l\leq l-2
\end{array} \right. \nonumber\\
&-&\left\{ \begin{array}{ll}
 p+1 & {\rm for}\; N-l\leq l-2 \\
 p+2l-N-1 & {\rm for}\; l-1\leq N-l+1\leq l+p
\end{array} \right. \nonumber\\
&=&\left\{ \begin{array}{ll}
 N-l-p & {\rm for}\; N-l-p<l \\
 l-1 & {\rm for}\; N-l-p\geq l
\end{array} \right.
\end{eqnarray}
where we used again that $p\leq N-l$. Hence (\ref{H_equ}) is proven.

\end{document}